\documentclass[useAMS,usenatbib,usegraphicx]{mn2e}

\def\kms{$\rm km\, s^{-1}$}
\def\cm3{$\rm cm^{-3}$}

\def\Vs{$\rm V_{s}$}
\def\n0{$\rm n_{0}$}
\def\B0{$\rm B_{0}$}

\def\Fh{$\rm F_{H}$}

\def\Ha{H$\alpha$}

\def\Fh{$\rm F_{h}$}

\def\mum{$\mu$m}
\def\Lx{L$_X$~}
\def\Fx{F$_X$}
\def\LIR{L$_{IR}$~}
\def\L12{L$_{12\mu m}$~}
\def\F12{F$_{12\mu m}$~}
\def\agr{a$_{gr}$}

\title[The Infrared - X-ray continuum correlation in AGN]{The Infrared -
X-ray continuum correlation in Active Galactic Nuclei}
\author[M. Contini, S.M. Viegas, and P.E. Campos]{M.
Contini$^{1}$\thanks{E-mail:
contini@ccsg.tau.ac.il}, S.M. Viegas$^{2}$ and P.E. Campos$^{2}$ \\
$^{1}$School of Physics and Astronomy, Tel Aviv University, Tel Aviv
69978, Israel\\
$^{2}$ Instituto de Astronomia, Geof\'{i}sica e Ci\^encias
Atmosf\'ericas, USP, Rua do Mat\~{a}o, 1226 05508-900  S\~{a}o Paulo,
Brazil\\}
\begin{document}

\date{Accepted . Received ; In original form }

\pagerange{\pageref{firstpage}--\pageref{lastpage}} \pubyear{2003}

\maketitle

\label{firstpage}

\begin{abstract}

The correlation between the soft X-ray and near infrared emission
from AGN  is analysed  using composite models. We 
find new evidences for differences
in ranges of parameters which characterize the NLR of Seyfert galaxies
and LINERs. Soft X-rays  show less variability, so they are 
better fitted for this kind of analysis.
In our models soft X-ray are emitted in the post shock region of clouds
with relatively high shock velocities \Vs $>$ 250 \kms. Consequently
dust emission peaks in the mid-IR.  On the other
hand, in the  photoionized zone, dust is at lower temperature
and  usually does not contribute to the mid-IR emission.
The results are sensible enough to  allow the same modelling method
to be applied to different types of AGN.
We found that  shock velocities are between 300 and 1000 \kms, 
with  the NLR of
low-luminosity AGN and Seyfert 2 showing lower velocities than
Seyfert 1 galaxies.
The intensity of the ionizing radiation flux at the Lyman 
limit from the central
source is low for LINERS and low-luminosity AGN 
(log \Fh = 9 to 10),
increasing towards Sy2 (log \Fh $\sim$ 11) and  Sy1
(11 $\leq$ \Fh $\leq $12).
Results obtained by modelling the Einstein and the ROSAT samples
of galaxies are in full agreement.
Dust-to-gas ratios by number are $\geq$ 10$^{-14}$ in LINERs and LLAGN,
 between 10$^{-15}$ and  3 10$^{-13}$ in Sy1 and up 
to  5. 10$^{-13}$ in Sy2.
In order to fit the infrared and X-ray continua, an
$\eta$ factor is defined,  which accounts for the emitting area
of the cloud. 
If the infrared emission is due to bremsstrahlung  and
comes from the same cloud producing the soft X-rays, the  $\eta$  values
obtained from both emissions must be the same.
Therefore, if ($\eta)_{IR} $ $<$ ($\eta)_{soft-X}$
there  must be   a strong contribution of soft X-rays from the active centre.
From the  $\eta$ values,   we expect to identify the objects 
that could present strong  variability. 

\end{abstract}


\section{Introduction}

The main source of ionization and heating of the emission-line
region  in active galactic nuclei (AGN) is photoionization by
radiation from the active centre (AC). Nevertheless, the contribution
of shock waves, generated by the radial motion of the clouds or
associated to jets, to the emission originated in 
the narrow line region (NLR) of Seyfert 1 (Sy1),
Seyfert 2 (Sy2) galaxies, and  LINERs was  pointed
out, for example, by Contini \& Viegas (2001), 
Contini (1997), and Viegas \& dal Pino (1992), and more recently
by  authors analysing the soft X-ray spectrum of AGN 
(e.g.  Iwasawa et al. 2003).
The relative importance of shocks and
radiation can be roughly  evaluated in the different objects, because
these processes act in different domains.
In fact,  shocks heat the gas within the clouds to high temperatures in 
the downstream  region, so
they can easily explain the observed high ionization lines 
(Rodriguez-Ardila et al. 2002) and  soft X-ray emission 
(Contini, Prieto \& Viegas 2002, and references therein).

Indeed, if soft X-rays are emitted  by bremsstrahlung from the NLR
clouds, the gas temperature must be at least  of the order  10$^6$ K.
Such  temperatures are obtained in the post-shock region if the  shock
velocity is higher than  250 \kms.
The FWHM line profiles in the NLR of Seyfert 2 and Seyfert 1 galaxies
indicate that the cloud velocities  can  well be beyond  this range.
Modelling the emission-line and continuum spectra of  some 
Seyfert 2 galaxies (e.g. NGC 5252, Circinus,   NGC 4151) indicate
the presence of very high shock velocities which can provide a
thermal origin for the observed soft X-rays.

Based on early observations of different galaxies by EXOSAT,
the nature and variability of soft X-rays was  investigated,
revealing  a low and long term variability compatible
with  bremsstrahlung emission from a hot gas. 
For instance, Pounds et al. (1986) showed 
the  existence of a soft X-ray spectral component,
distinct from the main (nuclear) X-ray emission 
of NGC 4151 by a lack of variability.
This component is shown to be consistent
with thermal emission from a hot, confining medium in the region of the
narrow emission-line clouds.
Perhaps NGC 4151 is  an exception to the general rule that soft X-ray
emission from AGNs is variable,  because most of its soft
X-ray flux is absorbed by a high intrinsic column density, so that its
faint,
diffuse X-ray flux can be detected. This effect should not
appear in most objects.

The correlation between the soft X-ray and  infrared luminosities
in AGN has been studied by several authors  and is controversial.
One important issue  being the source of variability in  the various
frequency ranges. In particular,  the correlations of \Lx (0.5-4.5 keV) 
versus \LIR ~for the IRAS 12\mum, 25\mum , 60\mum, and 100 $\mu$m bands
have been examined by  Green et al. (1992).
They claim that  \Lx ~versus  L$_{25\mu m}$ ~and \Lx versus
~L$_{100\mu m}$ resemble very closely the \Lx
versus L$_{60\mu m}$ correlation and that the plot  of \Lx ~versus \L12
~is considerably better defined,
yet, just as in the L$_{60 \mu m}$ plot,  "Seyfert galaxies still show no
significant correlation" (Green et al., 1992, Fig. 3)

On the other hand, 
the correlation  between the 10\mum ~and   the hard X-ray emission 
showed opposite results (Carleton et al. 1987  and Giuiricin et al. 1994) and
was reanalysed by Giuricin, Mardirossian,  \& Mezzetti (1995). They obtained
a weak correlation with a  harder (2 to 10 keV) X-ray emission,
mainly induced by Seyfert 2 data. They also found
that the 10\mum ~emission correlates well with the 12\mum 
 ~and 25\mum, and less well with the far-infrared emission 
at 60\mum ~and 100\mum. These authors favor thermal emission as 
the origin of the mid-infrared emission of Seyfert galaxies.

Dust is  present in the  NLR clouds and re-emission  by grains may
dominate the IR domain. If the intense UV-optical central 
radiation  heating the nuclear dust 
has an X-ray counterpart, a correlation between the infrared 
and  X-ray emission should be expected (Barvainis 1990). 
Because in shocked zones  the dust grains are
usually heated to higher temperatures than in photoionized 
regions, the contribution from the shocked
NLR clouds probably dominate the near infrared emission.  A
contribution  to the IR from bremsstrahlung emission from a cooler gas
is not excluded and would result in a correlation between the
the soft X-ray and the near IR continua. Nevertheless, 
in the former case the mutual heating of dust and gas (Viegas \& Contini
1994) can also lead to a correlation between those  two continua. 
Therefore, in both cases a correlation is expected. Notice that 
the contribution  of a dusty torus  to the IR emission suggested by the AGN
unified model (Antonucci \& Miller 1985) 
will give poor or no correlation with
the soft X-ray emission  
as pointed out by Green et al.(1992) 
(see also Contini, Viegas, \& Prieto 2003 and references therein).

A major problem concerning correlations between X-ray and IR emission
is X-ray variability. It is well established that soft X-rays from
low luminosity AGN (LLAGN) and LINERs show less and/or longer
variability than the hard X-rays  (Halderson, Moran, \& Filippenko 2001,
Terashima, Ho \& Ptak 2000). Following Green et al. (1992),
we will,  therefore, analyse the possible correlation between  mid-IR 
(\L12) and X-rays for these galaxies, which present a bolometric luminosity
that can be several orders of magnitude lower than the brightest AGN.

Notice that 10 \mum ~falls in the absorption trough of the spectral energy
distribution  due to silicate  grains. On the other hand, 
PAH emission bands dominate the  infrared SED of galaxies
at 6.2, 7.7, 8.6, 11.3, and 12.7\mum ~(Spoon et al 2002).
However, PAH grains are small compared with silicate grains
and easily sputtered in AGN, where shock  velocities are high enough
($\geq$ 300 \kms).  Therefore, we refer in this paper 
to the 12\mum ~emission in the mid-IR.

It was found in previous  investigations on
the Seyfert galaxies,  NGC 5252 (Contini, Prieto, \& Viegas 1998a),
Circinus (Contini, Prieto, \&
Viegas 1998b) and  NGC 4151 (Contini, Viegas, \& Prieto  2002), that
the critical shock velocity (\Vs) to produce  dust emission in the mid-IR 
is of the order of 500 \kms ~(Viegas \& Contini 1994),  
and that the critical  \Vs ~for  soft X-ray emission,
fitting  most of our sample for Seyfert 2 galaxies, is about 1000 \kms
(Contini \& Viegas  2000).
However, the dust-to-gas ratio, d/g, may change from galaxy to
galaxy, as well as in the different regions of a single galaxy.
Therefore a scattering of the infrared data  for  a sample of galaxies 
is expected.

The FWHM of the line profiles is  lower in LINERs ($<$ 300 \kms)
than in Seyfert galaxies (300 to 500 \kms). 
Notice, however, that broad \Ha ~lines
have been observed from LINERs (Ho, Filippenko, \& Sargent 1997), 
indicating that clouds with
higher velocities may also contribute to the spectra.
Therefore, in  these objects, characterized by a low ionizing flux from
the AC, shocks are relatively important and may contribute
to both the near-IR and soft X-rays emission. The observed data
will be confronted to model results obtained by numerical simulations
with the SUMA code (Viegas \& Contini 1994,
Contini \& Viegas 2001), which accounts for both the shock and
photoionization effects in the NLR clouds.

Once tested for the low-luminosity objects,
in the second part of this paper  we will  apply
the same kind of analysis  to Seyfert  galaxies.
We will restrict the data to the soft X-ray part
of the spectrum, where the variability is not as strong
as in the hard X-rays. From the comparison  with model results
we expect to  select those galaxies where
the contribution to soft-X rays from the
AC prevails, indicating in this way
the objects of the sample which should show a higher degree
of variability.

The spectral energy distribution (SED) of dust and gas emissions
 obtained from the models,
as well as  the corresponding temperatures of the dust grains
are discussed in Sect. 2. The \Lx - \L12 ~correlation for LLAGN and LINERs
is investigated in  Sect. 3, while
that for Seyfert galaxies is discussed in Sect. 4.
Concluding remarks appear in Sect. 5.

\section{Gas and dust emissions}

\subsection{The model}

\begin{figure}
\includegraphics[width=78mm]{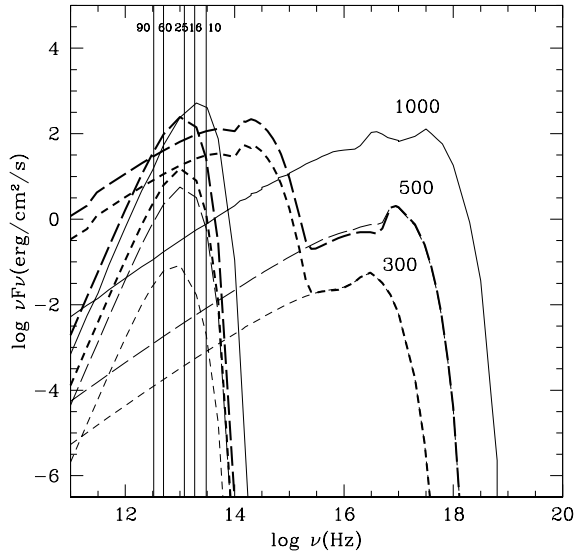}
\caption
  {
The SED of the continua emitted by bremsstrahlung and by dust 
from the clouds. Long-dashed lines refer to \Vs=500 \kms, log\Fh=12;
short dashed lines to \Vs=300 \kms, log\Fh=11. Solid lines refer to
\Vs=1000 \kms. Thick lines represent RD models, thin lines the SD models.
The vertical lines correspond to the main IR observed wavelengths.}

\end{figure}

In our model the  nuclear clouds  are moving outward from the
AC. A shock front appears at the cloud  outer edge, while
radiation from the AC reaches the inner edge. 
The physical conditions of a emitting cloud due
to the coupled effect of shock and photoionization by
a radiation source are obtained  with the SUMA code
assuming a plan-parallel symmetry. The calculations start at 
the shocked edge.
In the first run, only the shock effect is accounted for,
and the physical conditions are obtained for several ($\leq$ 300)
slabs before reaching the opposite
edge. In this first run
the results correspond to a shock-dominated model. Iterative runs
alternatively start  at the photoionized and shocked edges, accounting
for both effects, until convergence of the physical conditions
in the cloud is achieved.

The input parameters for a single-cloud model
are the shock velocity, \Vs, the preshock density,
\n0, the preshock magnetic field, \B0, the ionizing radiation spectrum,
the chemical abundances,
the dust-to-gas ratio by number, d/g,
and the geometric thickness of the clouds, D.  A power-law is adopted 
for the central ionizing radiation  spectrum reaching the cloud,
characterized by 
the flux \Fh, at the Lyman limit (in units of cm$^{-2}$s$^{-1}$ eV$^{-1}$)
and the power law index $\alpha$  (\Fh $\propto \nu^{-\alpha}$).
For all the models,  \B0 = $10^{-4}$ gauss, $\alpha_{UV}$ = 1.5
and $\alpha_X$ = 0.4, 
and cosmic abundances (Allen 1973) are adopted.
The model is either radiation dominated (RD) or shock dominated (SD) depending on
the intensity of the ionizing flux and on the shock velocity
(Viegas-Aldrovandi \& Contini 1989). For SD models \Fh = 0.

The gas entering the shock front is thermalized and heated.
 Depending on the shock velocity, a high temperature zone 
can be produced where soft X-rays are generated by bremsstrahlung. 
A lower  temperature zone may produce near infrared 
emission by the same mechanism.
The temperature  of the photoionized zone is never too high
($<$ 4. $\times$ 10$^4$ K), thus  its bremsstrahlung emission is mainly
in the IR-optical range, depending on the intensity of the ionizing
radiation. Thermal emission from dust grains could also produce 
near IR emission, as long as the dust temperature is high enough
as discussed below.

Before discussing the temperature
of the grains, let us illustrate
the SED of the continuum accounting for
both gas bremsstrahlung and dust re-emission calculated by a composite 
model (Contini \& Viegas 2000).
The SED shown in Fig. 1 correspond to single-cloud models with various cloud
velocities and
ionizing radiation intensities, assuming a geometric thickness
of the cloud D = $10^{19}$ cm and dust-to-gas ratio d/g =10$^{-14}$ by
number (or 4. 10$^{-4}$ by mass,
adopting silicate grains).
Following the results
obtained in our previous papers, models with high velocities (1000 \kms)
are SD, because the effect of the shock prevails 
on photoionization from the AC.
The results shown in Fig. 1 will help to understand the soft X-ray
emission radiation from clouds in  the physical conditions 
suitable to those in the NLR of AGN and in LINERs.

\subsection{The temperature of the dust grains}

\begin{table}
\centering
\caption{Model results : the temperature of grains}
\begin{tabular}{lll l l llll} \\ \hline

log \Fh & \Vs & \n0 & \agr  & T$_{gr}$ (SD)& T$_{gr}$ (RD) \\
- & (\kms) & (\cm3)& ($\mu$m) & (K) & (K) \\ \hline\\
- & 1000 & 1000 & 0.5 & 318. & -  \\
11 & 300 & 300 & 0.5 & 154.& 55.2  \\
- & 1000 & 1000 & 0.2 & 147. & -  \\
11 & 300 & 300 & 0.2 & 165. & 55.8  \\
- & 1000 & 1000 & 0.01 & - & -  \\
11 & 300 & 300 &  0.01 & 168.& 69.7 \\
12.7 & 300 & 300 & 0.01 & 168. & 112.\\
\hline\\
\end{tabular}
\label{tab1}
\end{table}

The  dust temperatures  calculated by models with \Vs = 300 and 1000 \kms
~are  presented in Table 1,  for both 
SD side and  the   RD side of the the single-clouds.
We assume that grains are present inside the whole cloud
before entering the shock front and all the dust grains
have initially the same size \agr. The grain radius may change due to
sputtering.  A dust-to-gas ratio d/g =  10$^{-14}$ by number 
is adopted.

In the photoionized zone, sputtering is not important
and the grain temperature
increases with decreasing grain radius  and increasing
\Fh. 
The behavior is rather different in the shocked zone because of
sputtering throughout the shock front.
The higher \Vs ~the  stronger the sputtering and the higher the
grain temperature, as long as the grains survive sputtering.
So, larger grains, which survive sputtering
through high velocity shock fronts, show a  distribution
of different radius throughout the cloud. 
They reach higher temperatures and
the peak of dust thermal emission shifts toward higher frequencies.
This peak can be high depending on  the volume of
the dusty region.  When the grain initial radius is small
and the shock velocity high the
dust grains are destroyed by sputtering, so dust cannot survive
in the SD side of the cloud (see Contini, Viegas, \& Prieto 2003).
Moreover there is  a mutual heating between gas and dust (see Viegas
\& Contini 1994). 

It can be seen that   T$_{gr}$
in the RD zone is always lower than in the
SD zone, even for small grains (\agr=0.01 $\mu$m) and
high \Fh~ (5 $10^{12}$ $\rm cm^{-2} s^{-1} eV^{-1}$).
On the other hand, dust is destroyed  by sputtering
if the initial grain radius is low
and \Vs is high, as exemplified by the model with  
\agr = 0.01 $\mu$m  \Vs = 1000 \kms.

{\it In brief,
if  the mid-infrared emission of AGN, particularly 
the 12$ \mu$ ~emission, is thermal emission due to grains,
the grains must be heated  in a shocked zone.}

\section{Low luminosity AGN}

\begin{figure}
\includegraphics[width=78mm]{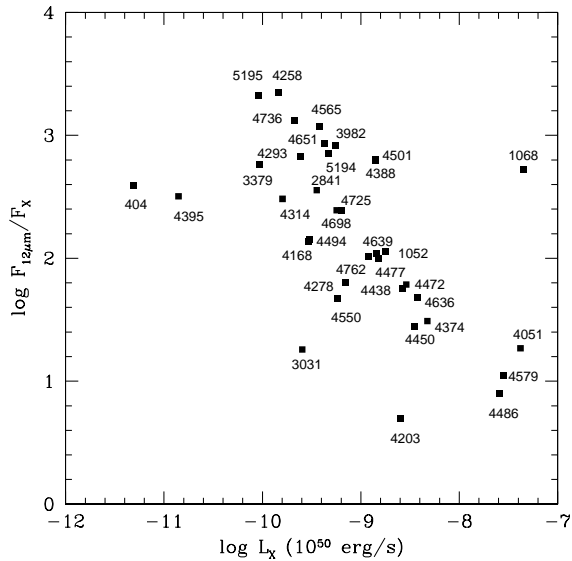}
\caption
  {
The ratio of the 12 $\mu$ luminosity to  the soft X-ray luminosity
 versus the soft X-ray luminosity
for the low-luminosity active galaxies of the  Halderson et al (2001)
sample.}

\end{figure}

The correlation between the near IR and the soft X-rays for  LLAGN is
analysed using  36 nearby objects from the Halderson et al. (2001) sample.
Most of the X-rays data come from observations with ROSAT while the
infrared data from IRAS. Some of the objects are also included in the
Terashima et al. (2002) sample of LLAGN observed with ASCA.
Short term variability ( $\leq$ 1 day) was found in NGC 3031,
NGC 4258 and NGC 5033. Long-term variability is observed
in NGC 3031, NGC 4258 and NGC 4579, while
no significant variability was found
in NGC 404, NGC 4111, NGC 4438, NGC 4579 and NGC 4639.

The correlation between the 12 \mum ~and soft X-rays for the LLAGN
is shown in Fig. 2, where we plot the ratio of the IR
and X-rays observed fluxes versus the X-ray luminosity. The
choice of the vertical axis is explained below, where
we compare  the observational data with the
results of numerical simulations.

\subsection{Modelling the sample}

\begin{figure}
\includegraphics[width=78mm]{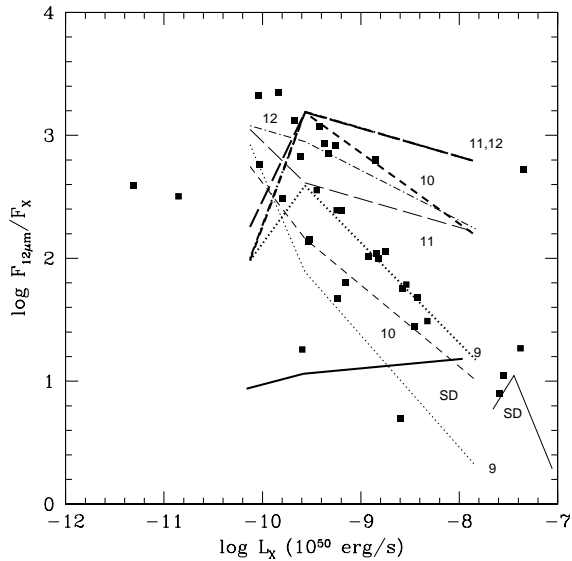}
\caption
  {
The comparison of model calculations with the observational data.
The results of the models with different values of \Vs and
the same log \Fh are joined by dotted (log \Fh = 9), short-dashed 
log \Fh = 10),
long-dashed ( log\Fh=11) and dash-dotted (log\Fh = 12) lines.
Solid lines correspond to shock dominated (SD) models calculated with
\Fh = 0
Thin lines refer to bremsstrahlung emission at 12 \mum, while thick
lines refer to
dust emission at 12 \mum. The values of \Vs increases from left to right.}

\end{figure}

The single-cloud model provides the continuum and emission-line fluxes
 calculated at the cloud. Thus, in order to obtain the
luminosities, the model results must be multiplied by a
factor $\eta$ accounting for the emitting area. 
Assuming that the single cloud conditions correspond
to the gas around the central source, the emitting area
can be related to  the
distance of the cloud to the central source and to the covering
factor (see, for instance, Contini \& Viegas 2000).
Plotting a flux ratio versus a luminosity, as in Fig. 2,
this factor will affect only the horizontal axis,
and the match of the models to the data would be  easier,
if IR and X-ray emission come from the same cloud and there
is a negligible contribution of X-rays from the central source.

Since we assume  that the soft X-rays come from the shocked
edge of the NLR clouds, the post-shocked zone must reach temperatures
higher than  10$^5$ K and  a minimum shock velocity of  300 \kms
~is required. Based on previous works, a grid of
models with  \Vs=300 \kms and \n0=300 \cm3,
\Vs=500 \kms and \n0=300 \cm3, \Vs=700 \kms and \n0=500 \cm3,
and   log \Fh = 9,10,11,12 was obtained for clouds with a
dust-to-gas ratio of 10$^{-14}$, a geometrical width of  D=10$^{19}$ cm,
a pre-shock magnetic field of \B0=10$^{-4}$ gauss and initial grain radius of
0.2  $\mu$m. 
Notice that log \Fh = 12 is an upper limit since
a    low radiation flux from the AC is characteristic of LLAGN.

Model results and observational data are compared in Figure 3.
The lines join results of models with different values of ~\Vs
but the same \Fh.

The observations of the galaxies NGC 4486, NGC 4579, and NGC 4051
are better explained by SD models (thin solid line joining the
results  for
\Vs = 900, 1000, and 1100 \kms ~and d/g = 10$^{-13}$).
Only the bremsstrahlung is shown in this case,
because at these high velocities the grains are mostly destroyed by
sputtering.
The  \F12/\Fx ~ratios for SD models  calculated  with \Vs= 300, 500, and
700 \kms
~and d/g=10$^{-14}$ are  too low. Higher values
(thick solid line) are obtained assuming d/g = 2$\times 
10^{-13}$.

The observed correlation is reproduced by  models calculated with log
\Fh=9 and 10. However, it is difficult to distinguish between bremsstrahlung and
thermal emission by dust.

Assuming  all clouds at the same distance (r)
from the center with the covering factor equal to unity,
the best general fit to the observations, which reproduces
the slope of the correlation,  is obtained with clouds at r = 15.45 pc.

\subsection{Results for individual galaxies}

\begin{table}
\centering
\caption{The characteristics of LLAGN}
\begin{tabular}{lll l l cc} \\ \hline
\  NGC     & class  &	\Vs  & \n0  &  log \Fh  & d/g	       & r \\
\	   &	    &  (\kms)& (\cm3)&    -	& (10$^{-14}$) & (pc) \\
\hline
\   404 & L2	 &  500   & 300  & 11	     & 1.   & 1.5 \\
\   1052& L1.9   &  500-700& 300-500 & 10    & 1.   & 15.5 \\
\   1068& S1.9   &  700   & 500      & 11    & 1.   & 23.6 \\
\   2841& L2	 &  500   & 300      & 9     & 1.   & 15.5 \\
\   3031& S1.5   &  700   & 500      & SD    & 20.  & 2.4  \\
\   3379& L2	 &  300   & 300      & 9     & br.  & 15.5 \\
\   3982& S1.9   &  500-700& 300-500 & 10-12 & 1.-br.& 15.5 \\
\   4051& S1.2   &  1000   & 500     & SD    & br.   & 15.5 \\
\   4168& S1.9   &  500    & 300     & 10    & br.   & 15.5 \\
\   4203& L1.9   &  500-700& 300-500 & 9     & br.   & $>$ 15.5 \\
\   4258& S1.9   &  500    & 300     & 12    & 1.    & 8.9 \\
\   4278& L1.9   &  500-700& 300-500 & 10    & br.   & 15.5 \\
\   4293& L2	 &  300-500& 300     & 10    & 1.    & 15.5\\
\   4314& L2	 &  300-500& 300     & 10    & br.   & 15.5\\
\   4374& L2	 &  500-700& 300-500 & 9     & 1.    & 15.5\\
\   4388& S1.9   &  500-700& 300-500 & 10    & 1.    & 15.5\\
\   4395& S1.8   &  500-700& 300-500 & 11    & 1.    & 1.55\\
\   4438& L1.9   &  500-700& 300-500 & 9     & 1.    & 15.5\\
\   4450& L1.9   &  500-700& 300-500 & 10    & br.   & 15.5 \\
\   4472& S2	 &  500-700& 300-500 & 9     & 1.    & 15.5\\
\   4477& S2	 &  500-700& 300-500 & 9     & 1.    & 15.5\\
\   4486& L2	 &  900    & 500     & SD    & br.   & 8.9\\
\   4494& L2	 &  500    & 300     & 10    & br.   & 15.5 \\
\   4501& S2	 &  500-700& 300-500 & 10    & 1.    & 15.5 \\
\   4550& L2	 & $>$ 500    & $>$300     & 9 & br. & 15.5 \\
\   4565& S1.9   & $>$ 500   & $>$ 300&  10-12 & 1.  & 15.5-8.9\\
\   4579& S1.9   & 1000& 500 &         SD    & br.  & 8.9\\
\   4636& L1.9   & $<$700& $<$ 500    & 9     & 1.   & 15.5 \\
\   4639& S1.0   & 500-700 & 300-500  & 9     & 1.   & 15.5\\
\   4651& L2	 & 500     & 300      & 12    &  br. & 15.5\\
\   4698& S2	 & $>$ 500 & $>$ 300  & 9     & 1.   & 15.5 \\
\   4725& S2	 & $>$ 500 & $>$ 300  & 9     & 1.   & 15.5 \\
\   4736& L2	 & $>$500  & $>$ 300  & 12    & 1.   & 8.9 \\
\   4762& L2	 & 500-700 & 300-500  & 9     & 1.   & 15.5\\
\   5194& S2	 & $>$ 500& $>$ 300 & 12      & br   &  15.5 \\
\   5195& L2	 & 500     & 300    & 12      & 1.   & 8.9  \\
\hline
\end{tabular}
\label{tab2}
\end{table}

\begin{figure}
\includegraphics[width=78mm]{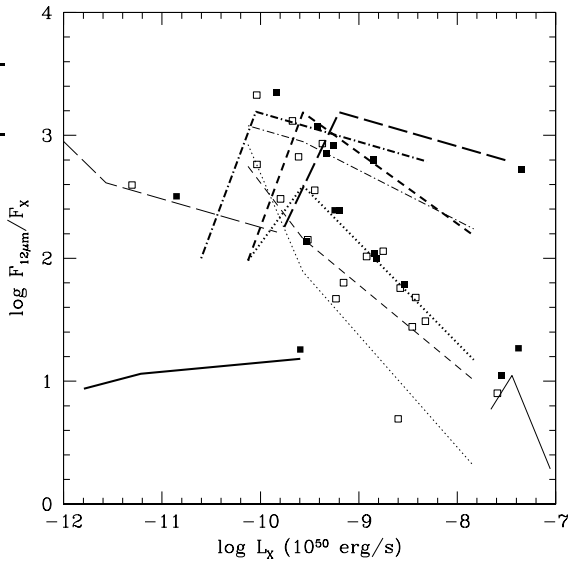}
\caption
  {
The best fit of the data by model results (see text).}

\end{figure}

Single-cloud models for the galaxies in the Halderson et al. 
sample are chosen based on the two data:
the infrared luminosity at 12 \mum ~and that in the soft X-rays (0.5-2.4 
keV).
In the latest years point-to-point observations  of
a single objects permit to obtain   information
about  the distribution of velocities, densities,  intensity of the
radiation from the AC, and  dust-to-gas ratios throughout the galaxy.
By modelling the soft X-ray continuum, however, we  select
the clouds with shock velocities high enough to heat the
gas in the post-shock region to  high temperatures.
The  modelling of these clouds permits to evaluate  roughly their
distances from
the AC.  In Table 2, for each galaxy of the sample
the average characteristics of the emitting clouds reproducing the
observed correlation are listed.

The galaxies of the Halderson et al. sample are separated in LINERs
(empty squares) and LLAGN (full squares) in Fig. 4.
Some of the curves representing the results from models shown in Fig. 3
appear
horizontally shifted in Fig. 4 in order to reach most of the objects.
This shift corresponds to an $\eta$ factor different from that
used in Fig. 3.
As seen in Fig. 4  most of the LINERs are
fitted by low \Fh ~(log \Fh=9-10), confirming that  low \Fh ~are a
characteristic of these objects
(Contini 1997). The clouds in NGC 404 and NGC 4395  that are closer to
the AC, are reached by a rather high flux
(log \Fh=11). Our results suggest that NGC 4168 should be
classified as a LINER. NGC 3031  has  shock dominated clouds close
to the nucleus (r=2.4 pc) heated by high velocity shock  (\Vs=700 \kms).
NGC 1068, on the other hand, shows IR emission by dust from clouds
with d/g=10$^{-14}$ and \Vs=700 \kms, located at 23.6 pc from the center.
Clouds in NGC 4486 and NGC 4579  have the highest shock velocities (\Vs
$\sim$ 1000 \kms)
and are located in the nuclear region (r=8.9 pc).
Notice that NGC 1068, which has an H$\alpha$ luminosity greater than
the limit for LLAGN (10$^{40}$ ergs s$^{-1}$), does not follow the
general trend (see Fig.2), since high velocity clouds at 
large distance from the center are required to account 
for the observed L$_X$ (Table 2).
NGC 4051 is a warm absorber and soft X-ray emission  has been
explained by a more complete model
(Contini \& Viegas 1999).

\subsection{Composite radio emission}

\begin{figure}
\includegraphics[width=71mm]{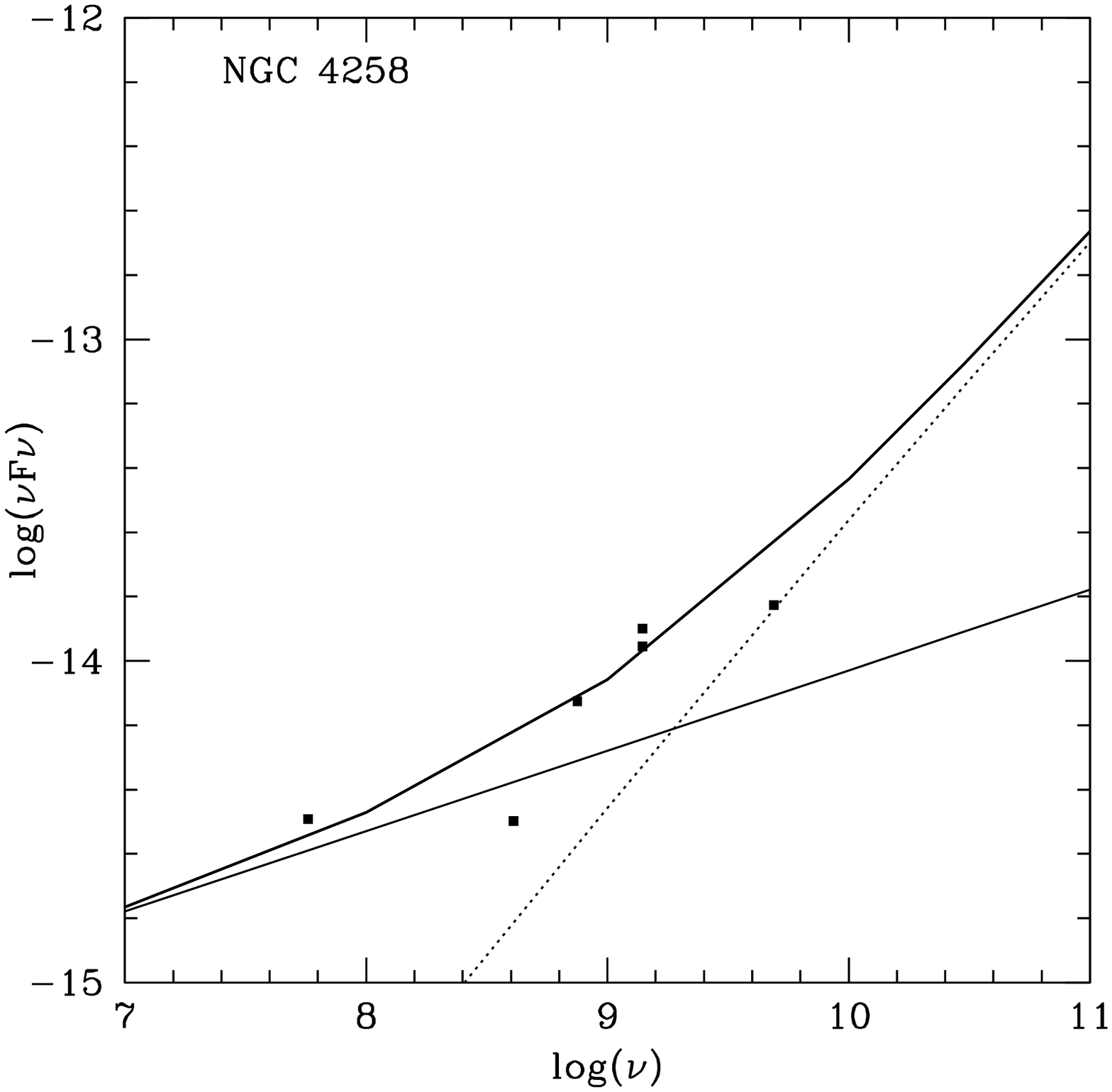}
\includegraphics[width=71mm]{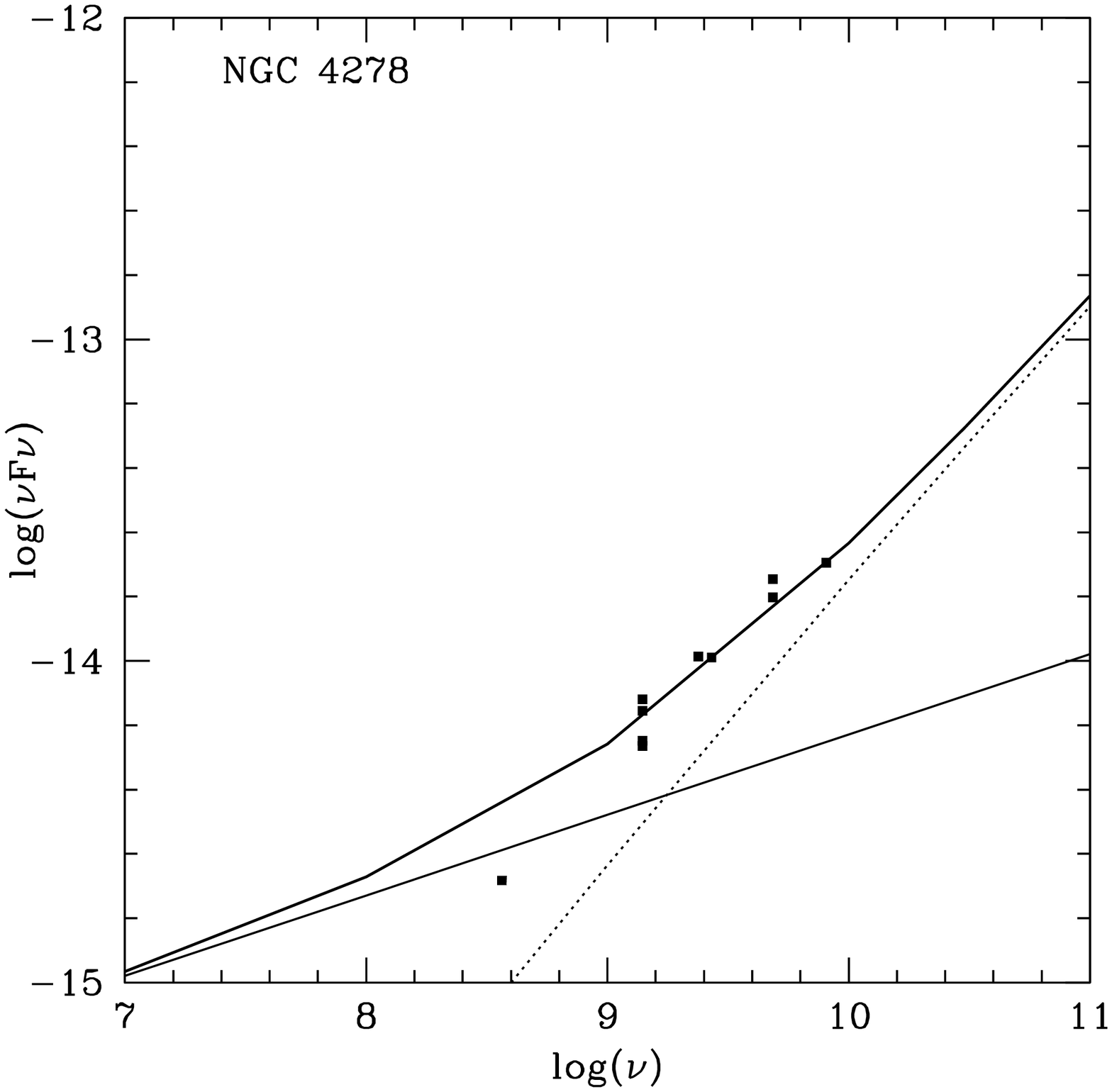}
\includegraphics[width=71mm]{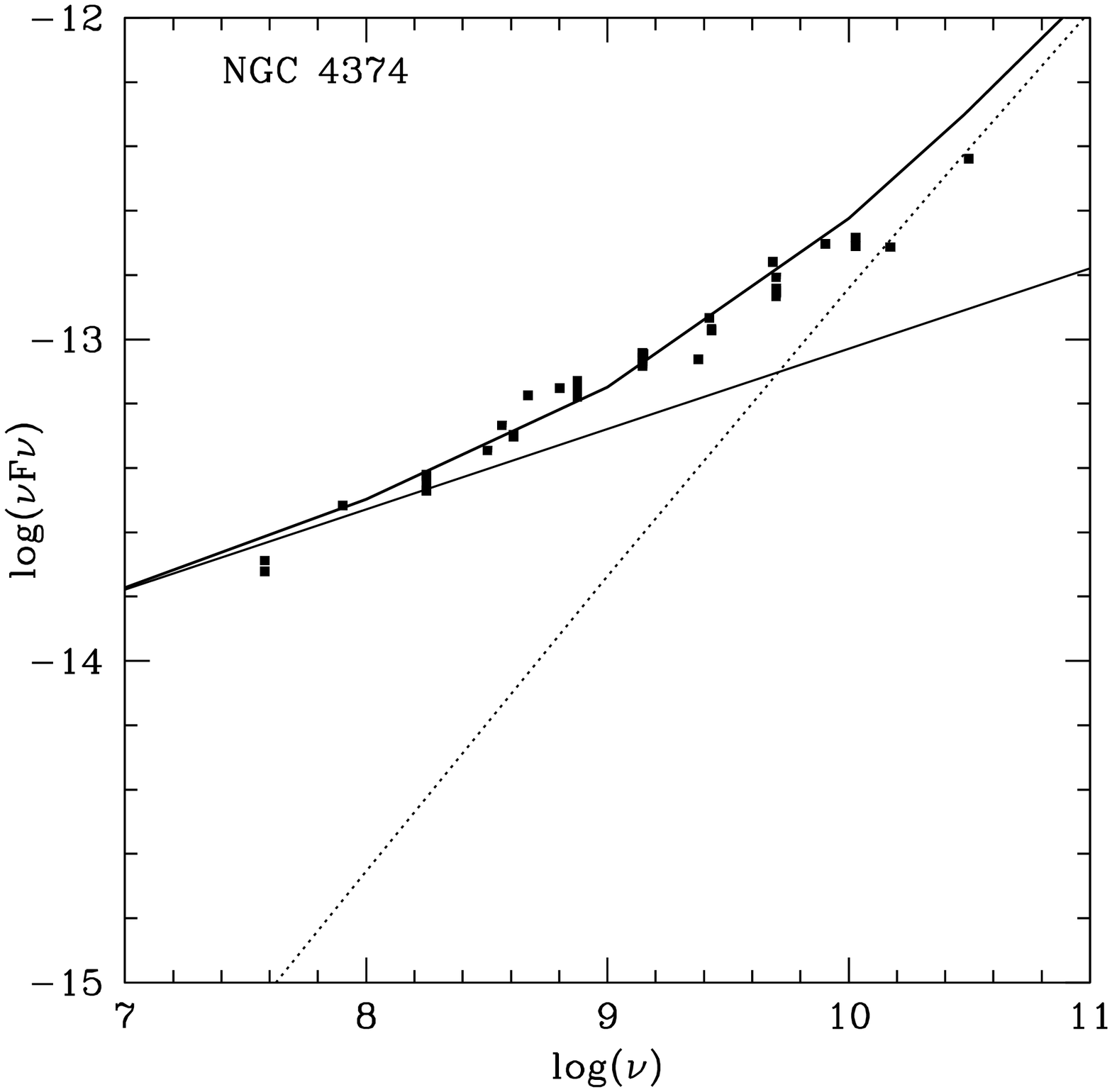}
\caption
  {
The  calculated radio continuum
for NGC 4374, NGC 4278, and NGC 4258. The two components
are shown: the power-law due to Fermi mechanism (thin solid line) and the
bremsstrahlung from a cool gas (dotted line)
The thick solid line corresponds to the sum of the two components.}

\end{figure}

Before concluding the analysis of the
 LLAGN, we would like to strengthen the  fact that shocks and
the ionizing radiation 
from the AC act together  heating and ionizing the gas also in low
luminosity objects as illustrated by the radio continuum.
In fact, the trend of the SED in the radio domain is obtained 
by summing up
the contribution from the  synchrotron radiation created by 
the Fermi mechanism at the shock front and  the bremsstrahlung from 
the cooler cloud zones. 

A few examples appear in Fig. 5, showing
model results and observational data for the galaxies NGC 4374 (top
panel, NGC 4278 (middle panel), and NGC 4258 (bottom panel). 
The model input parameters are listed in Table 2.The
observational data were found 
in the NASA Extragalactic Data (NED)) and come from 
Laing \& Peakock 1980, Kuhr et al. 1981, Kellermann et al 1969, Gower
et al. 1967,
Wright \& Otrupcek 1990), Douglas et al. 1996, Ekers 1969, Large et al.
1981, Heeschen \& Wade 1964, White \& Becker 1992, Waldram et al. 1996,
Becker et al. 1995, Condon et al. 1983, Dressel \& Condon, 1978, Becker
et al. 1991, Gregory \& Condon 1991, Israel \& Mahoney 1990, 
Ficarra et al. 1985.

\section{The \Lx - \L12 correlation for Seyfert galaxies}

\begin{figure}
\includegraphics[width=78mm]{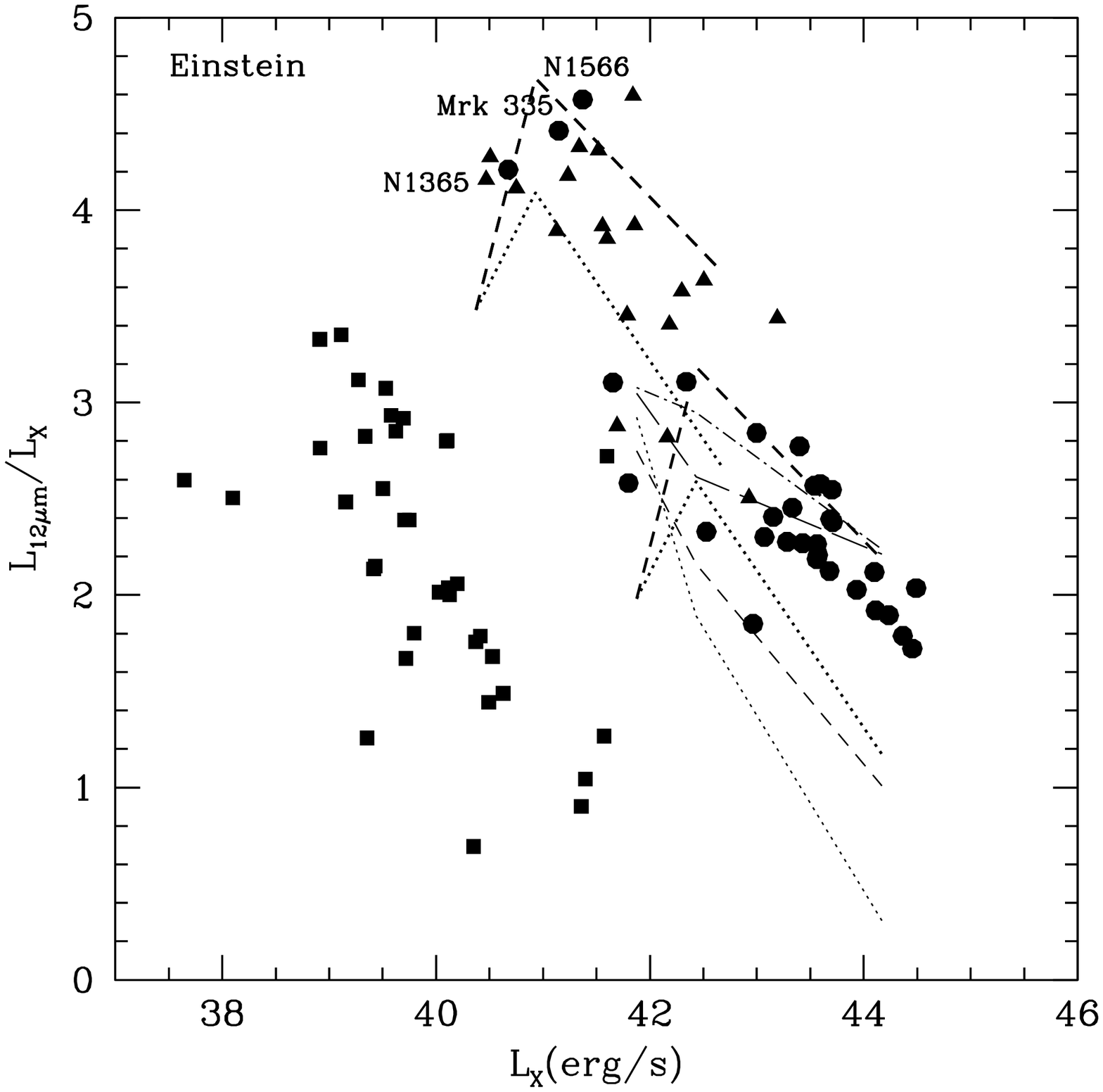}
\includegraphics[width=78mm]{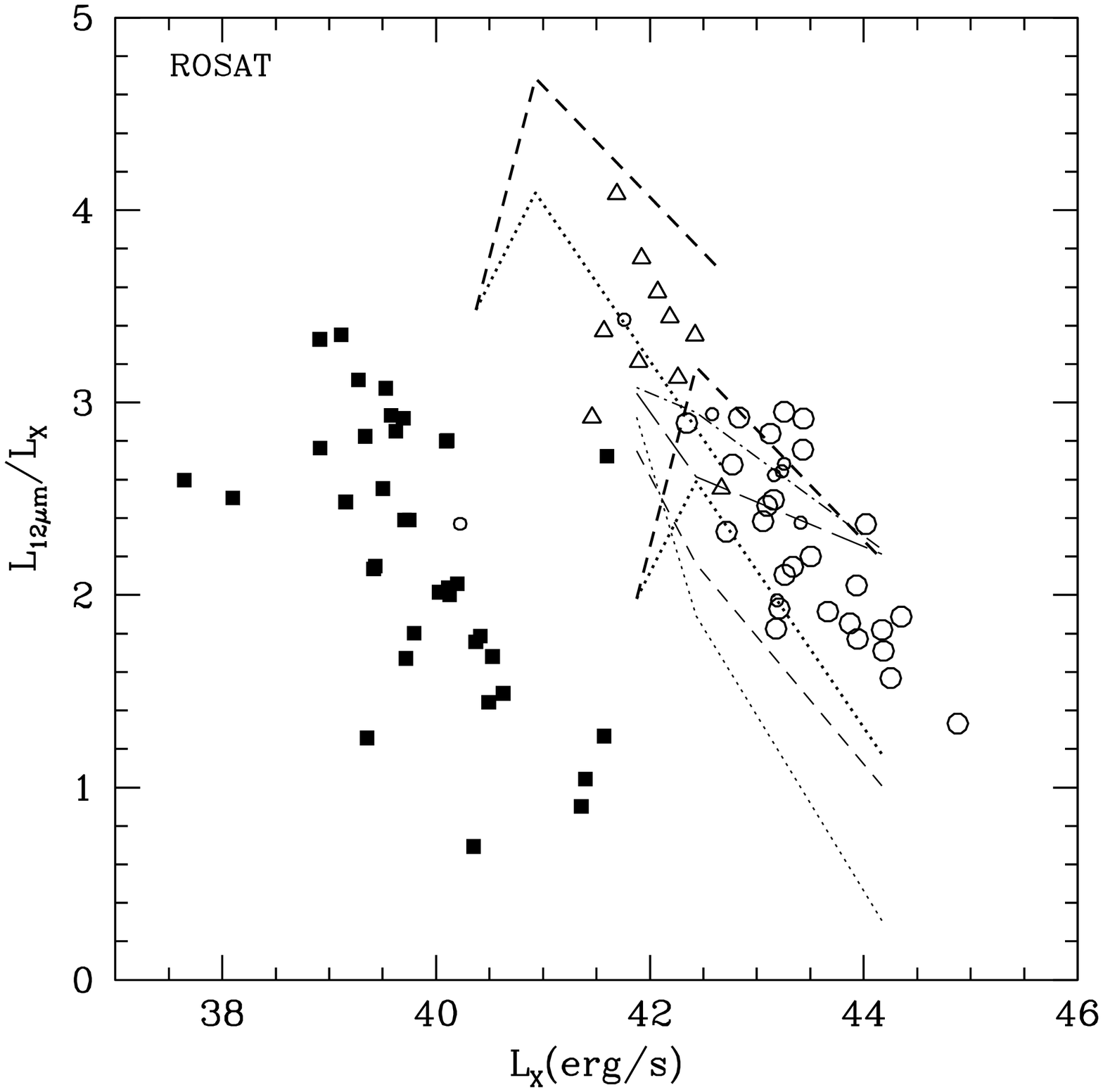}
\caption{
Top panel : \L12/\Lx versus \Lx for Sy1 (filled circles) and Sy2
(filled triangles) galaxies using Einstein data.
Bottom panel :
\L12/\Lx versus \Lx for Sy1 (open circles) and Sy2 (open triangles) 
galaxies using  ROSAT data.
In both panels, filled squares represent the LLAGN
and the lines correspond to model results with the same
notation used in Fig. 3.}

\end{figure}

\begin{table*}
\centering
\caption {The logarithm of the  $\eta$ factors using the Einstein data}

\begin{tabular}{ll   c c ccc lccc} \\ \hline\\
type & model &X-ray$^1$ & IR(br.) & IR(dust) & r (pc)& d/g
(10$^{-14}$)&line type\\ \hline\\
Sy1 & RD  & 43.8 & 44.2  & -&747. &-& thick long-dashed \\
Sy1&  SD & 42.6&- & 46.0 &187.6&251.& thin short-dashed\\
Sy1 & RD  & 44.1&- & 45.2&1055.&1.26 & thick short-dashed\\
Sy2 & SD&40.1 & - & 44.5&10.5&2512.& thin short-dashed\\
Sy2 & SD& 40.1 & - & 45.3& 10.5&1.58(4)&thin dotted\\
Sy2 & RD & 43.3 & - & 44.8 & 420.&3.16& thick short-dashed\\
Sy2 & RD & 43.3&  - & 45.3 &420.&10.0& thick dotted\\
\hline\\
\end{tabular}

$^1$ Einstein data

\label{tab3}
\end{table*}

\begin{table*}
\centering
\caption {The logarithm of the $\eta$ factors using the  ROSAT data}

\begin{tabular}{ll   c c ccc lccc} \\ \hline\\
type & model &X-ray$^1$  & IR(br.) & IR(dust)& r (pc)& d/g
(10$^{-14}$)&line type\\ \hline\\
Sy1 & RD  & 44.1 & 44.1 & -    &1055.&- & thick long-dashed\\
Sy1&  SD & 42.5 & - & 46.2& 167.2& 501.&thin short-dashed\\
Sy1 & RD  & 43.9 & - & 44.9 &838.&1. & thick short-dashed\\
Sy2 & SD&40.1 & - & 44.3 &10.5&1585.& thin short-dashed\\
Sy2 & SD& 40.1 &-& 45.1&10.5 & 1.(4)& thin dotted\\
Sy2 & RD & 43.2 & -& 44.6 &374. & 2.5 &thick short-dashed\\
Sy2 & RD &43.3 & - & 45.4&420. &12.6& thick dotted\\
\hline\\
\end{tabular}

$^1$ ROSAT data
\label{tab4}
\end{table*}

In this section the correlation between the soft X-ray  luminosity (\Lx)
and the 12 \mum ~luminosity (\L12) is  investigated for a relatively large
sample of Seyfert galaxies which includes objects
with higher luminosities  relative to  those of LINERS and LLAGNs studied
in the previous sections.

We have collected a sample of 31 Sy1 and 19 Sy2  observed with
the Einstein (0.5 - 4 keV band)  and IRAS satellites. Another sample
includes 26 Sy1, 8  Sy1.5 and 10 Sy2 observed with ROSAT (0.5 - 2.4 keV).
The Einstein data come from Green, Anderson \& Ward (1992), the ROSAT
data were taken from Moran, Halpern \& Helfand (1996). The IRAS data
come from the NED.
The luminosities are calculated assuming a Hubble constant
equal to 50 \kms  Mpc$^{-1}$.

The correlation between  \L12/\Lx 
 and  \Lx for Sy1 and Sy2 galaxies are shown in Fig. 6 using
Einstein (top panel) and ROSAT data (bottom panel), respectively.
For comparison, the LLAGN data (Helderson et al. 2001) are also plotted.
As expected the LLAGn are fainter in X-rays than the Seyfert galaxies.
As expected, NGC 1068, included in the Halderson et al. sample, is
located in the Seyfert galaxy zone.

The separation between the two classes of objects is more clearly marked
for ROSAT data.
The \L12/\Lx ratios are roughly  of the same order. These ratios
depend  mainly on the dust-to-gas ratio  and on the physical conditions
in the NLR, indicating that the conditions 
are similar in the two classes of galaxies.

The  same series of models
(same notations) applied to  the LLAGN in Fig. 3
are used to roughly model the Seyfert galaxies.
In order to fit the data for the  Einstein sample,
 we adopted an $\eta$  corresponding to an
average distance of the clouds from the AC  of $\sim$ 460 pc for Sy1 and
$\sim$  90 pc for Sy2.
This corresponds to shifting the model  ensemble to
the right hand side in the diagram. However, to find the best fit for Sy2  
we had to
shift the models upwards by a factor of 30. The vertical-axis corresponds
to the \L12 / \Lx ratio, where \L12 represents dust emission.
Therefore,
the best fit corresponds to dust-to-gas ratios higher 
by a factor of 30 (d/g = 3 $\times 10^{-13}$)
and  already suggests that the NLR of Sy2 is dustier than that of Sy1.
The same parameters used for the Einstein sample are adopted
to fit the ROSAT sample (Fig. 6 bottom).
Notice that Mrk 335, NGC 1365, and NGC 1566 from the Einstein sample,
showing low \Lx, should be classified as Sy2 rather than Sy1.

The fit shown in Fig. 6 is reasonable,
although more refined modelling is necessary.
This will be done in the next section, where   the NLR 
physical conditions, as well as the evaluation of the distances
of the Sy1 and Sy2 emitting clouds to the center,  are  discussed.
For this, the \L12 versus \Lx correlation is used for both
Sy1 and Sy2 galaxies (Fig. 7 and 8, respectively, corresponding
to the Einstein and ROSAT data).

\subsection {Modelling the correlation}

Single-cloud models that better reproduce the \LIR - \Lx 
correlation shown in Figs. 7 and 8 are chosen from a grid of models. 
All models are calculated with d/g = 10$^{-15}$ by number,
\B0= 10$^{-4}$ gauss, and \n0 = 300, 500, and 1000 \cm3 for models with
\Vs = 300, 500, and 1000 \kms, respectively (cf Fig. 1).
Usually  in the NLR the decreasing radiation flux with the
distance to the AC is followed by a decreasing velocity field.
Moreover, we adopted a low  dust-to-gas ratio as a lower limit, in order to find out the
whole range of d/g values for both  Sy1 and Sy2.

The theoretical results are represented
by lines, labelled by the input parameters \Vs ~for 
SD models (thin lines),
and the pair (\Vs,log \Fh) in the case of RD models (thick lines).
Short dashed lines refer to dust emission in the IR, while long dashed 
lines to IR bremsstrahlung. 

Fitting of the observed correlation with our
models is not so simple.
First, recall that  models correspond to single clouds 
(see Sect. 3.2).
Thus, the comparison between the theoretical
and observational data can only give a rough indication of
the physical conditions of the gas  and of the velocity field
of the clouds emitting soft X-rays.
Second, as already commented above (Sect. 3.1),
in order to derive the luminosities, the fluxes
calculated by the models must
be multiplied by a factor ($\eta$)
which  represents the fraction
of the observed luminosity coming from a specific type of
cloud, characterized by \Vs, \n0 and \Fh.  If the infrared
emission is due to dust, this factor must also account for the
differences of the dust-to-gas ratio assumed in the calculations
and the real value for each galactic nucleus.
A higher d/g value leads to an enhancement of the L$_{IR}$ emission.
Thus, we expect that the  $\eta$ factor is not constant but
may have  different values for different galaxies. 
It could be empirically obtained
by fitting the observed emission-line and continuum spectra
with a multi-cloud model, as already presented
for Circinus and NGC 5252 (Contini et al. 1988 b,a).

Since our goal is to analyse the \L12 - \Lx correlation
we try to obtain a rough picture of the physical processes
in the active nuclei with a minimum number of input parameters.
First we constrain the models by considering the 
main trends shown by the data,
because these trends define which type of models must be adopted (RD ,
SD, IR from dust or IR from bremsstrahlung).
Then, we  shift the  model results in order to cover 
the largest number of data by changing the $\eta$ factor.

Recall that in the SED presented in Fig. 1, the high frequency
domain corresponds to the hot gas emission, i.e., even in radiation
dominated models it depends on the shock velocity.
So, the soft X-ray emission is due to bremsstrahlung in the shocked
zone of high velocity clouds while its infrared counterpart
can contribute to the mid- or near-IR emission. In this case, since
both continua come from the same cloud, the corresponding
$\eta$ should be the same.

The logarithm of $\eta$  for the bremsstrahlung 
and for dust emission, adopted for modelling the correlations presented
in Figs. 7 and 8, are  given in Tables 3 and 4 for Einstein and ROSAT 
data, respectively.
In the first column  of Tables 3 and 4 the type of the galaxy is given.
In the ROSAT sample, Sy1 includes the Sy1.5 data.
The model type (SD or RD) is indicated in the second column.
The logarithm of the  $\eta$ value which provides
the best fit of the data by the corresponding model
(see line type in column 8)
 are given in  columns 3, 4, and 5
for  \Lx, \L12 generated by bremsstrahlung, 
and \L12 ~by dust, respectively.
In columns 6 and 7 the calculated distance of 
the emitting clouds from the AC
(assuming a covering factor  equal to unity)
and the d/g ratio by number are given, respectively.
The values of d/g  by number which better  fit the data are calculated
from the difference between $\eta$ corresponding to IR(dust) and $\eta$ corresponding
to X-ray for  each type of galaxies in Tables 3 and 4.
The line type used to draw the curves corresponding
to the models (Figs. 7 and 8)  are listed in the last column.

\subsection{Seyfert 1 galaxies}

Two clear trends are present in Fig. 7  (top panel)
corresponding to  the Sy1 Einstein sample:
(a) Sy1 with \Lx $\leq$ 3 $\times$ 10$^{43}$ show a larger scatter of the
\L12 values; (b) the objects with higher  \Lx values are mainly
located between RD (thick short-dashed line) and SD ( thin short-dashed line)
models both with IR due to dust emission .
The results corresponding to \L12 due to bremsstrahlung
(thick long-dashed line) could eventually explain some Sy1 and Sy2 data.

Regarding the ROSAT data (Fig. 7, bottom panel),
most of the data  are also inside a sector
defined by RD models (the low edge) and SD models (upper edge),
as found  for the Einstein sample.
High \L12 ~Sy1 galaxies are better fitted by dust emission from SD models
with 300 $<$ \Vs $\leq$ 1000 \kms ~ (thin line), whereas those with low
\L12 would be dominated by dusty RD clouds with  \Fh=  10$^{11-12}$ (
$\rm cm^{-2} s^{-1} eV^{-1}$) and \Vs ~in the range
300 - 500 \kms.
Also for the ROSAT data, the line corresponding to the IR bremsstrahlung 
emission fits  some Sy1 and Sy2 data, but not the trend of the Sy1 complex.

\subsection{Seyfert 2 galaxies}

The sample of Sy2  is shown in Fig. 8  both
for Einstein (top panel) and ROSAT data (bottom panel).
In both  samples, the data are better
explained by RD or SD models with \L12  produced by dust thermal emission.
The notation is the same as in Fig. 7. Notice that thin 
short-dash lines refer to SD models and thick short-dash
to RD models with IR from dust, while dotted lines correspond
to the same models shifted  in the diagrams toward higher \LIR,
in order to explain the high IR luminosity of objects
which are rich in dust.

Except for four objects in the top panel and three in the
bottom panel, most of the data points are located between
two SD curves, characterized by log $\eta$ = 40.1 and 43.3
On the other hand, the objects on  right of the the
SD curves are close to the RD results (thick short-dash)
with log \Fh=11, \Vs ~between 300 and 500,
and a relatively low d/g (see Tables 3 and 4).

\begin{figure}
\includegraphics[width=78mm]{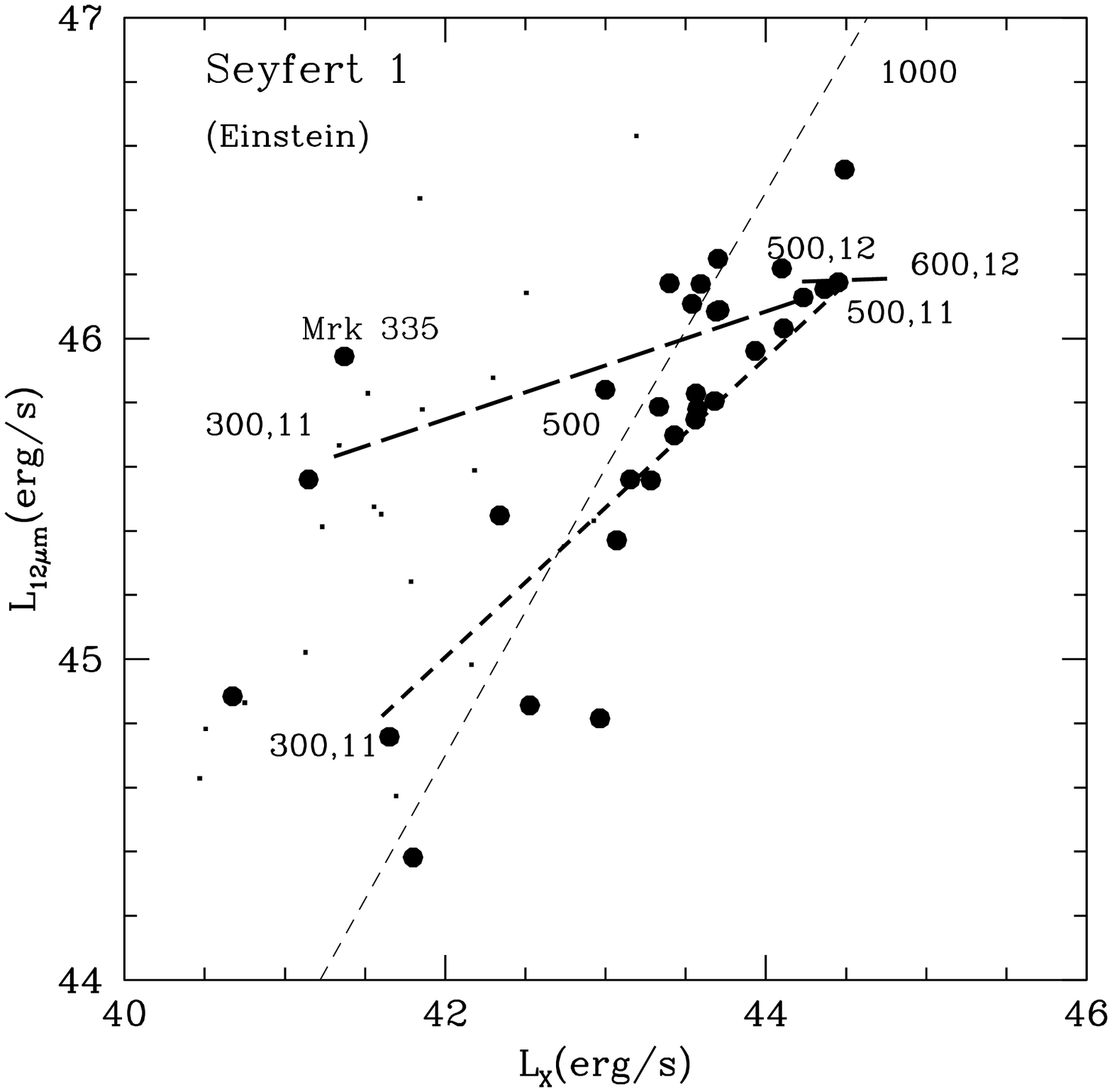}
\includegraphics[width=78mm]{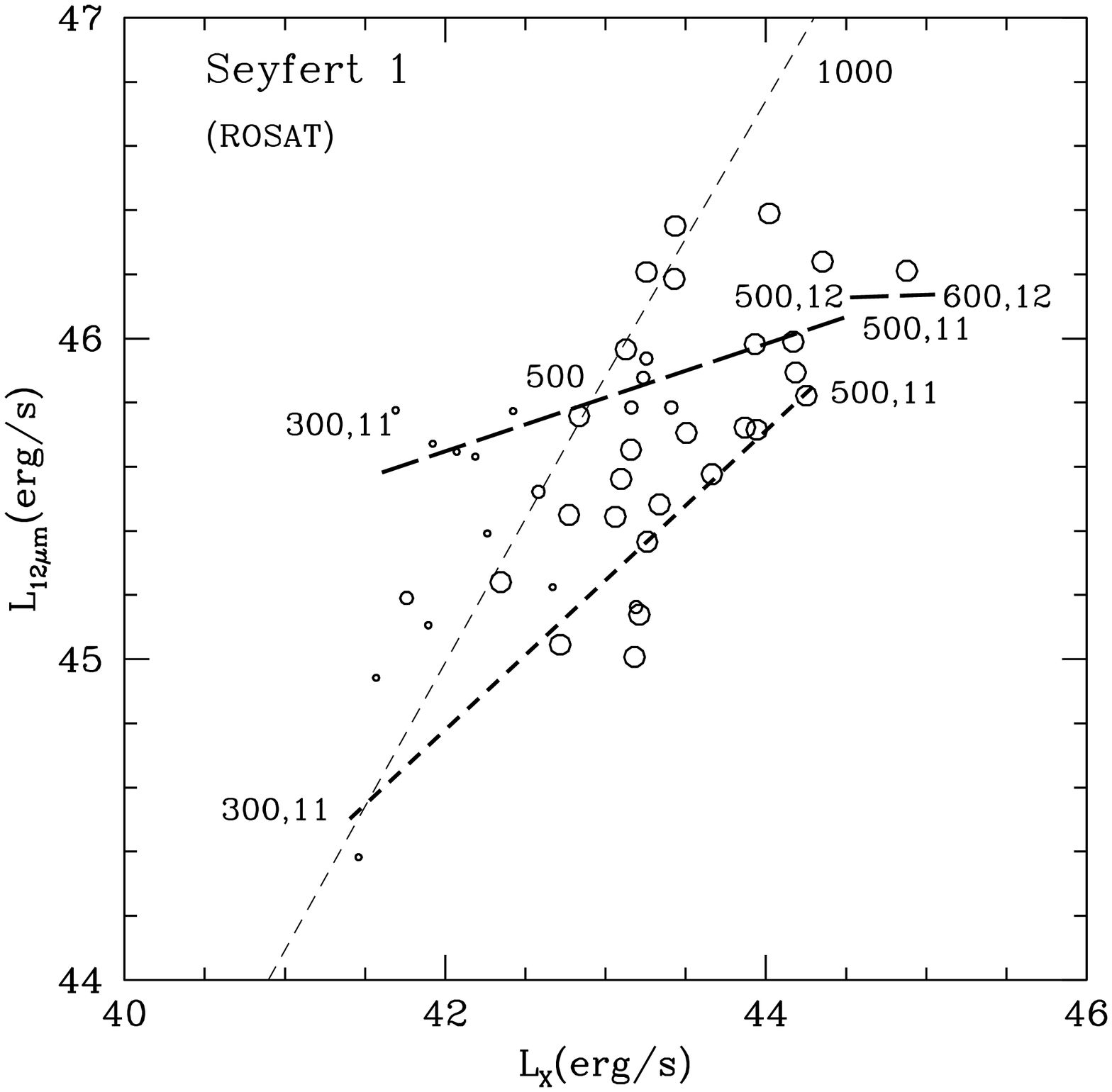}
\caption{
\L12 versus \Lx for Seyfert 1 galaxies (larger symbols).
Top panel refers  to the Einstein sample, while the bottom panel to ROSAT
sample.
The curves are labelled by the  \Vs and log \Fh
values for RD models and by  \Vs  for SD models. The
line types are  given in Tables 3 and 4 (last column).
For the Einstein data the notation is:  Sy1 (filled circles);
For the ROSAT data the  open circles represent  Sy1, and small open
circles the Sy 1.5. In
both diagrams,  tiny symbols  correspond to Sy2 data.}

\end{figure}

\begin{figure}

\includegraphics[width=78mm]{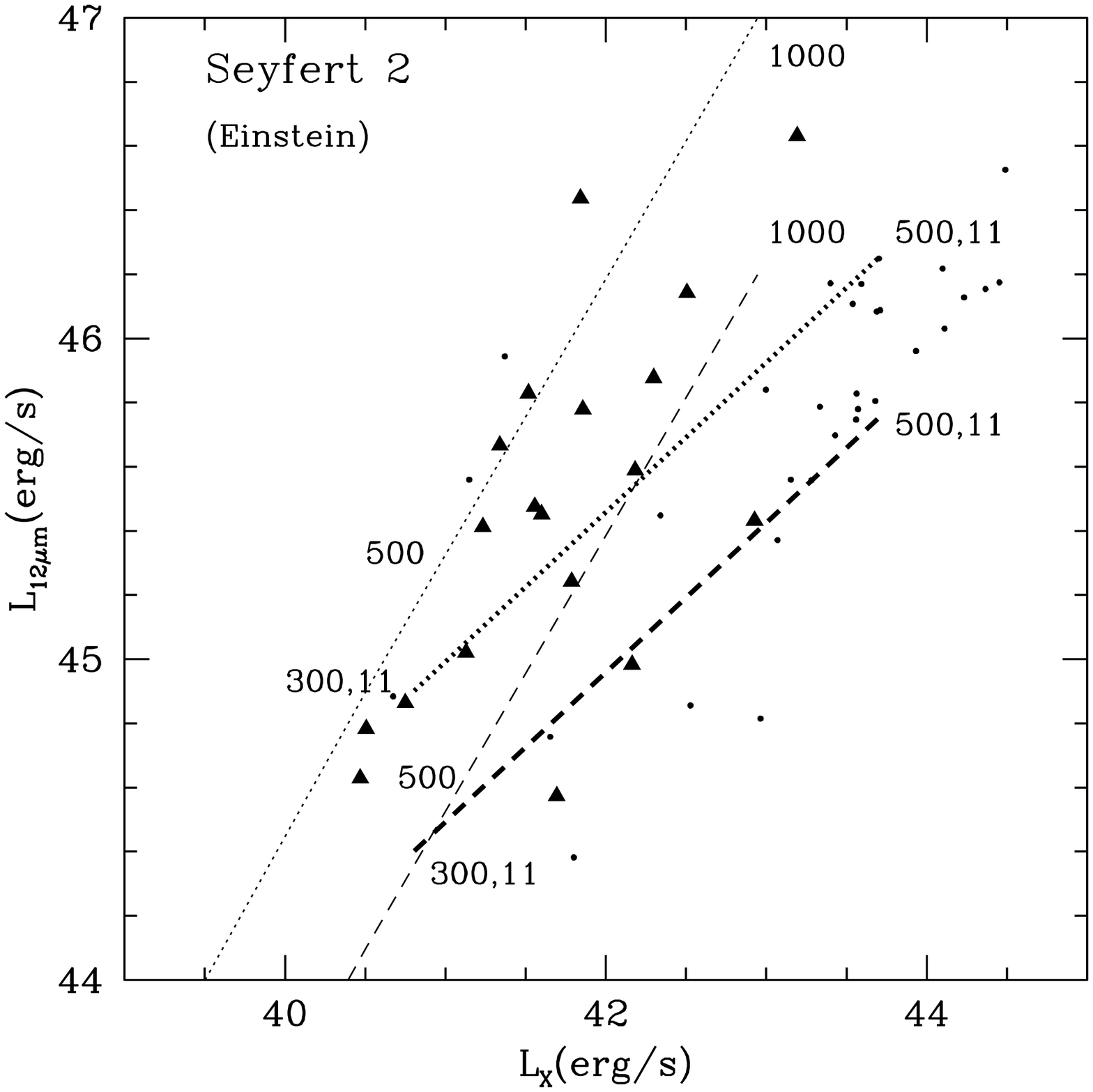}
\includegraphics[width=78mm]{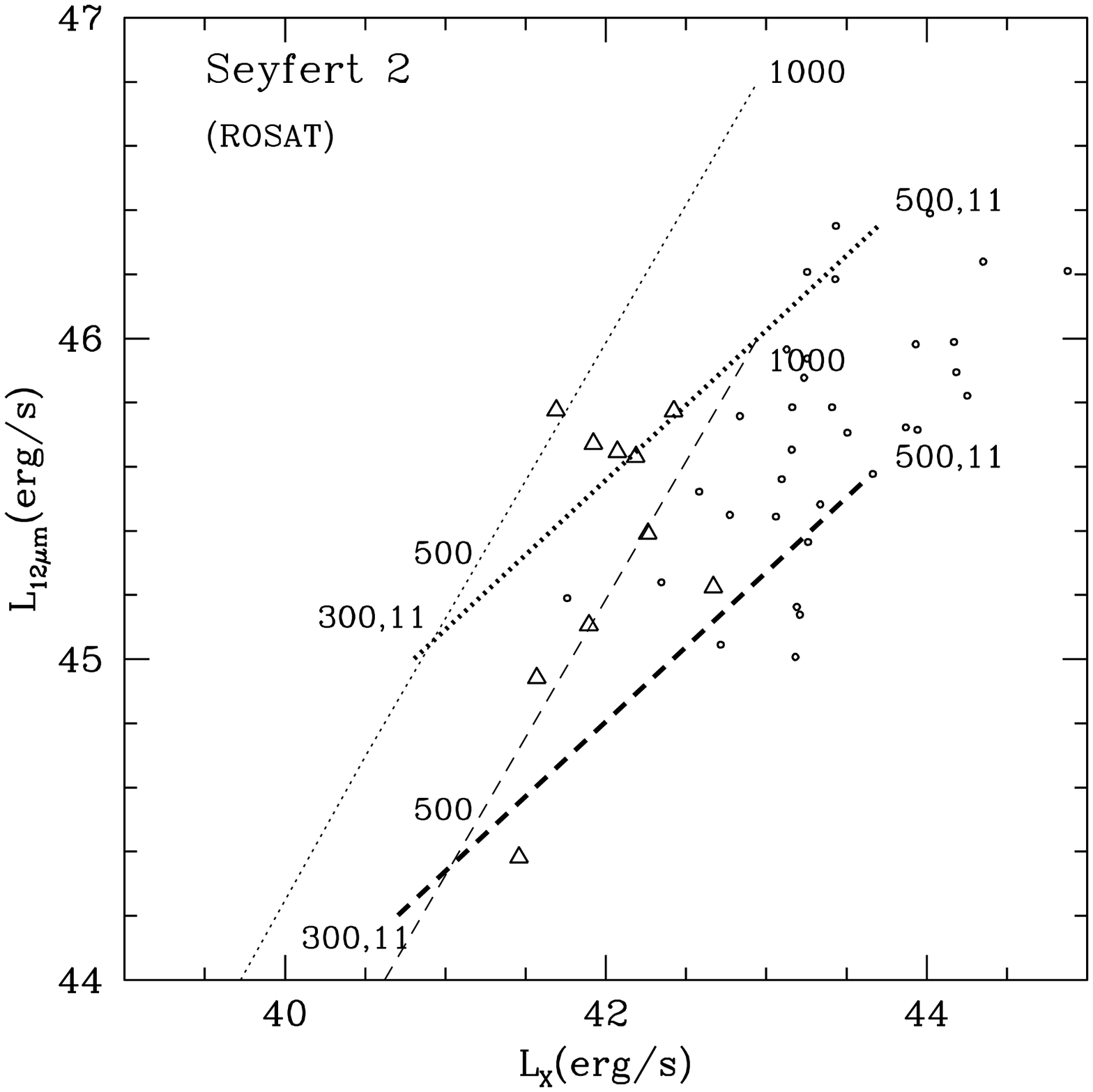}
\caption{
\L12 versus \Lx for Seyfert 2 galaxies (larger symbols).
Top panel refer  to the Einstein sample, while the bottom panel to ROSAT
sample. The curves are labelled by the  \Vs and log \Fh
values for the RD models and by  \Vs  for SD models. The
line types are  given in Tables 3 and 4 (last column).
Filled triangles represent the Einstein sample (top panel)
and open triangles (bottom panel) the ROSAT sample.
In both panels tiny symbols correspond to the Sy1 data.}

\end{figure}

\subsection{Discussion}

The first results of our investigation show that the X-ray and near-IR
correlation for Sy1 can be explained by the composite models of clouds
producing  both bremsstrahlung and dust emission. In the diagrams
these models cover most of the area where the AGN data points are
distributed.

In a photoionized zone, the dust cannot reach a temperature
high enough to contribute to the 12$\mu$m emission, whereas
a shocked zone may heat the dust to higher temperatures, as
shown in Table 1. Therefore, models accounting for the
shock are important.
Considering the whole sample for a given class (Sy1 or Sy2),
the contribution of RD and SD clouds to  the IR emission
coming from thermal emission by dust provide  the
best fit to the correlation for both Einstein and ROSAT data.

The values of the factor $\eta$
adopted for the models and  presented in Tables 3  and 4 give a
rough estimation of the distance of the emitting nebula
from the AC.
Notice, however, that the observed \Lx may  include a
contribution  from the central source, while \L12 depends on d/g.
So, the $\eta$ value corresponding to  \L12 due to bremsstrahlung,
which is related to the distance of the cloud to the AC,
can be lower than that corresponding to \Lx.

The ratio of $\eta$ for \Lx and \L12 will provide
an estimate of the \Lx emitted from the AC  relative to \Lx
emitted by bremsstrahlung.
For example, the observational data (Einstein sample) for the
Sy1 galaxy Mrk 335, could be  reproduced by IR from dust 
with a high d/g. However, it could be also reproduced
by IR produced by bremsstrahlung, with
log$(\eta)_{soft-X}$=43.7 and log$(\eta)_{IR}$=43.6.
 A difference
in the   $\eta$ values showing
log$(\eta)_{soft-X}$  higher than log$(\eta)_{IR}$
 indicates a  contribution from the AC
to \Lx. In this case, this galaxy  could show soft X-ray short term
variability.

The theoretical results shown in Figs. 7 and 8  correspond
to models with  d/g =$10^{-15}$. With a
lower d/g, \L12 dust will hardly prevail over \L12 bremsstrahlung,
therefore, also in the case of emission by dust we assume as
indicative of the nebula radius the lowest value of $\eta$.
The Sy2 galaxies  included between the thin short-dash and
dotted curves of Fig. 8  have a d/g  between 2.5 10$^{-12}$ and 1.6 
10$^{-11}$
(Einstein) and between 1.6 10$^{-12}$ and 10$^{-11}$ (ROSAT).
The results for the Sy1 galaxies  indicate
d/g   = 2.5 10$^{-13}$ - 1.3 10$^{-14}$ (Einstein)
and  5. 10$^{-13}$ - 10$^{-14}$ (ROSAT).

Regarding the objects with IR coming from dust re-radiation,
results obtained for the galaxies of  both Einstein and ROSAT
samples, we obtain from the values of $\eta$ listed in Tables 3 and 4
a characteristic distance of a few hundred pc to a few  Kpc
for the Sy1.
The distances for Sy2 galaxies  are
smaller going from a few pc  to about 500 pc.

In brief, the results obtained by the modelling the
\Lx - \L12 correlation  suggest that
both SD and RD clouds are present in Sy1 and Sy2.
Notice, however, that for Sy2 galaxies, \Vs ~is about 300 \kms and log \Fh
~is of the order of 11, while Sy1 have  higher \Vs ~(up to 600 \kms)
and higher \Fh ~(log \Fh up to  12).

Clouds closer to the nuclear region are shock dominated
because a large amount of dust
can prevent the ionizing radiation from the AC.
High velocity  (\Vs=1000 \kms) and dust rich  SD clouds are more likely
to be present in Sy2 than in Sy1, and  are closer to the center.
This indicates that  the emitting region in Sy2 is more compact
than in Sy1. However,
the maximum distance of a few  Kpc
obtained for Sy1 from the fit of the ROSAT sample  could
represent an upper limit  if the emitting gas  has both a high d/g
and a strong contribution to the X-ray coming from the active nucleus.
 Schmitt et al. (2003) found that both Seyfert types
have similar distributions of the NLR sizes, 
although it is expected  that Sy1 have in average smaller NLR sizes than Sy2
(Mulchaey et al 1996). Notice that Schmitt et al claim that there are "effects which could alter
both the measurements and models".

Finally, the results obtained  from the  ensemble of the Seyfert galaxies
reveal that the X-ray - near-IR  emitting regions of Sy1 and Sy2 have
indeed different characteristics, which are hardly  noticeable  when
modelling single objects.

\section{Concluding remarks}

In this paper we looked for new evidences for 
differences in  parameter ranges
characterizing the NLR of Sy1, Sy2, and LINERs (and LLAGN) by
modelling of the \Lx - \L12 correlation by composite models accounting
for dust emission.

Soft X-rays are considered because they show less dramatic variability.
In our models, soft X-ray are emitted in the post-shock
region of clouds with relatively high
shock velocities. Dust and gas are coupled crossing the shock front,
therefore,
dust emission from the  cloud region that emits the soft X-rays peaks in
the mid-IR.
Shocks are relatively important in modelling the LLAGN spectra, so, we
have started our
investigation  with the \LIR  - \Lx correlation  for LLAGN.
The results are sensitive enough to apply the same modelling method to
Seyfert galaxies.

We found that  shock velocities are between 300 and 1000 \kms for all
the objects, lower in LLAGN and Seyfert 2, higher 
in the NLR of Seyfert 1 galaxies.
The intensity of the flux from the AC is low 
for LINERs and LLAGN (log \Fh=9 - 10)
and increases towards Sy2 (log \Fh $\simeq$ 11), being higher for Sy1
(log \Fh $\leq $12).
Results obtained by modelling the Einstein sample 
and the ROSAT sample are in full agreement.
Dust-to-gas ratios by number are larger than 10$^{-14}$ 
(4. 10$^{-4}$ by mass) for LINERs and LLAGN,
between 10$^{-14}$  and  3 $\times$ 10$^{-13}$ (4. 10$^{-4}$ and 0.012
by mass) for Sy1, and  reach
2. $\times$ 10$^{-11}$ (0.8 by mass) for Sy2.

The soft X-ray variability of Seyfert galaxies can be
inferred from the $\eta$ values deduced from the models.
If ($\eta)_{IR} $ $<$ ($\eta)_{soft-X}$  there is a strong
contribution of soft X-rays from the AC to the observed value.
Therefore,  objects of the sample used in this paper which
could present strong variability can be identified.

\section*{Aknowledgments}

We are very grateful to  an unknown referee for
precious comments.
This paper is partially supported by
FAPESP (00/06695-0),CNPq (304077/77-1).

\bsp

\label{lastpage}

\end{document}